\documentclass[submission]{eptcs}
\usepackage{fixltx2e}

\usepackage[utf8]{inputenc}
\usepackage[T1]{fontenc}   
\usepackage{amsmath}
\usepackage{tikz,cite}
\usepackage{tabularx}
\usepackage{amssymb}
\usepackage{xspace}
\usepackage{amsfonts,stmaryrd}
\usepackage{alltt}
\usepackage{hyperref}
 \usepackage{graphicx}
\usepackage[export]{adjustbox}
 \usepackage{listings}
\usepackage{mathpazo}

\usepackage[english]{babel}

\title{Formal Safety and Security Assessment of an Avionic Architecture with Alloy}

\author{Julien Brunel
\institute{ONERA/DTIM \\ F-31055 Toulouse, France}
\email{julien.brunel@onera.fr}
\and 
Laurent Rioux \quad Stéphane Paul
\institute{Thales Research \& Technology\\ F-91767 Palaiseau, France}
\email{firstname.lastname@thalesgroup.com}
\and Anthony Faucogney \qquad Frédérique Vallée
\institute{All4Tec\\ F-53001 Laval, France}
\email{firstname.lastname@all4tec.net}
}
\def\titlerunning{Safety and Security Assessment of an Avionic Architecture with Alloy}
\def\authorrunning{J. Brunel,  A. Faucogney, S. Paul, L. Rioux \& F. Vallée}

\lstdefinelanguage{Alloy}
{morekeywords={abstract, all, and, as, assert, but, check, disj, else, enum,
  exactly, extends, fact, for, fun, iden, if, iff, implies, in, Int,
  int, let, lone, module, no, none, not, one, open, or, part, pred,
  run, seq, set, sig, some, sum, then, this, univ},
sensitive=true,
morecomment=[l][\itshape\color{gray}]{--},
morecomment=[l][\itshape\color{gray}]{//},
morecomment=[s][\itshape\color{gray}]{/*}{*/},
mathescape,
identifierstyle=\tt,
keywordstyle={\tt\bfseries},
tabsize=2, columns=fullflexible, 
literate=
{->}{{$\to$}}1
{&}{{\&}}1
{^}{{$\mspace{3mu}\widehat{\thinspace}\mspace{3mu}$}}1
{<}{$<$}1
{>}{$>$}1
{>=}{$\geq$ }2 
{=<}{$\leq$ }2 
{<:}{{$<\mspace{-2mu}:$ }}3
{:>}{{$:\mspace{-2mu}>$ }}3
{=>}{{$\Rightarrow$ }}2 
{+}{$+$}1
{++}{{$+\mspace{-8mu}+$ }}3
{\~}{$\sim$}1
{!=}{$\neq$}1
{*}{${}^{\ast}$}1 
{\#}{$\#$}1 
}

\begin{document}
\lstset{language=Alloy}

\maketitle

\begin{abstract}
 We propose an approach based on Alloy to formally model and assess a
 system architecture with respect to safety and security
 requirements. We illustrate this approach by considering as a case
 study an avionic system developed by Thales, which provides guidance
 to aircraft. We show how to define in Alloy a metamodel of avionic
 architectures with a focus on failure propagations. We then express
 the specific architecture of the case study in Alloy. Finally, we
 express and check properties that refer to the robustness of the
 architecture to failures and attacks.
\end{abstract}


\section{Introduction}
Nowadays, safety and security are commonly identified disciplines in
system and software engineering. However it often covers the same
meaning in the head of non-experts. Even if the end target may be
considered similar, \emph{i.e.}, “protect the system against
unspecified behaviors during operation”, the approaches proposed by
safety and security and their evaluation criteria are sometimes
equivalent but sometimes quite different or even opposite.




For example, at risk management level, on the one hand, a security
policy can require keeping a door closed to secure the data saved in a
room, and on the other hand, the safety policy can require keeping
this door open in case of fire. How is the relative importance of both
concerns managed? Does safety serve security or does security serve
safety or are they at the same level?  Another example at development
level: the safety and security disciplines are also different, handled
by different teams, which most of the time do not talk to each other
and where their work is based on different vocabulary and
standards. What would the common language be for making better
communication between safety and security teams?

Besides, the mindset of safety and security teams may be different,
even opposite when we focus on knowledge sharing. The first one tries
to share the most valuable knowledge in order to reduce impacts due to
systems failures; the second one tries to protect its security
mechanisms against publishing in order to reduce the potential
exploitation probability. How can safety and security team mindsets may
be harmonized and better coordinated?

A collaborative ITEA2 project named
MERgE\footnote{http://merge-project.eu} has been setup to identify the
discipline divergences and convergences to then demonstrate the
benefit of a co-engineering activity supported by a collaborative
modeling tool platform. In this article we focus on a solution
promoted by the project to perform early validation of systems with
respect to safety and security properties, taking benefit from
lightweight formal methods, and more specifically
Alloy~\cite{Jackson:06}.

This article is organized as follows. In Sect.~\ref{sec:LPV}, we
present the case study, a system that provides guidance to
aircraft. In Sect.~\ref{sec:eval-case-study}, we show how to model in
Alloy the architecture of the case study. Then, we express and check
the expected safety properties in Sect.~\ref{sec:assessm-safety-secur}
and the security properties in Sect.~\ref{sec:security-assessment}.



\section{Case study description}
\label{sec:LPV}

An approach is the last leg of an aircraft’s flight, before
landing. Approaches can be performed under Visual Flight Rules (VFR)
or under Instrument Flight Rules (IFR). VFR are a set of regulations
under which a pilot operates an aircraft in weather conditions clear
enough to allow the pilot to operate the aircraft with visual
reference to the ground, and by visually avoiding obstructions and
other aircraft. For the aircraft operating under IFR, an Instrument
Approach Procedure (IAP) defines a series of predetermined
instrument-supported manoeuvres from the beginning of the initial
approach to the landing (or to a point from which a landing may be
made visually). There are two main classifications for IAPs: precision
approaches (PA) and non-precision approaches (NPA). Precision
approaches utilize both lateral and vertical
information. Non-precision approaches provide lateral course
information only. Instrument approaches have traditionally been
supported by land-based navigation aids (i.e. the so-called
“navaids”), typically Localizers, VHF Omnidirectional Range (VOR)
beacons, Non-Directional Beacons (NDB), Distance Measuring Equipment
(DME), or Instrument Landing System (ILS).

Since the 1960s, Global Navigation Satellite Systems (GNSS) have been
experimented and deployed to bypass land-based equipment. GNSS is a
system of satellites that provide autonomous geo-spatial positioning
with global coverage. It allows small electronic receivers to
determine their location (longitude, latitude, and altitude) to high
precision (within a few metres) using time signals transmitted along a
line of sight by radio from satellites. Currently, the United States
NAVSTAR Global Positioning System (GPS) and the Russian GLONASS are
the only global operational GNSSs. The European Union's Galileo
positioning system is a GNSS in initial deployment phase, scheduled to
be fully operational by 2020. Localizer Performance with Vertical
guidance (LPV) is the highest precision GNSS aviation instrument
approach procedure currently available without specialized aircrew
training requirements. LPV is designed to provide 16 meter horizontal
accuracy and 20 meter vertical accuracy 95 percent of the time. LPV
landing minima are usually similar to those in an ILS.

\begin{figure}\begin{center}
\includegraphics[width=.99\textwidth]{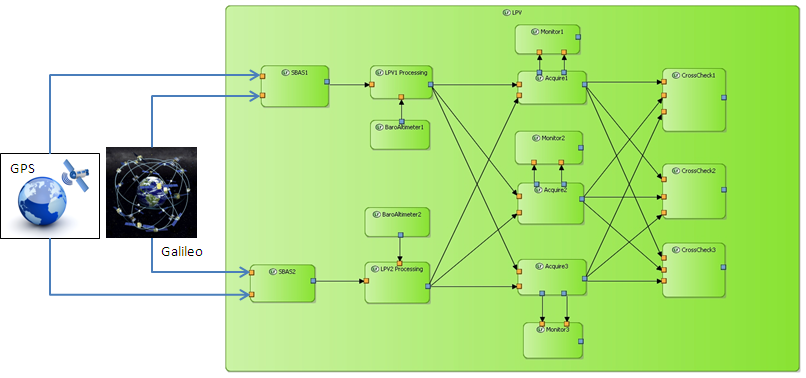}
\end{center}
\caption{\label{fig:LPV-archi}LPV architecture}
\end{figure}

This case-study is concerned by the architecting of a new Thales
Avionics aircraft embedded system designed to support an LPV
approach. Its architecture is represented by Fig.~\ref{fig:LPV-archi}.

We can summarize the behavior of this sub-system as follows. Both
satellites constellations (American GPS and GALILEO constellations)
send the signal to \texttt{SBAS} processing functions. After
correlations of both positions information, the \texttt{SBAS} sends
the plane position (lateral and vertical) to two occurrences of the
\texttt{LPV processing} function. The data produced by \texttt{LPV
  processing} functions are sent to three displays (three occurrences
of a function \texttt{Acquire}. In each display, a comparison of the
data received from \texttt{LPV1} and \texttt{LPV2} is performed. In
case of inconsistency, an alarm is triggered by a function
\texttt{Monitor}. The crew chooses which of the two LPV processing is
used by each display (function \lstinline{SelectSource}, not
represented in Fig~\ref{fig:LPV-archi}). Besides, each display
receives the data computed by the two other displays. Then, the
function \texttt{Crosscheck} compares the data of the current display
with the data of the two others and resets the current display in case
it differs from both other displays.

This initial architecture was designed taking into account a number of
safety requirements. Two of these requirements are recalled below.

\begin{tabular}{|l p{12.5cm}|}
\hline
&\textbf{Safety requirements:}\\
\emph{Safety 1} &\emph{Loss of LPV capability}.
No single failure must lead to the loss of LPV capability.
\\
\emph{Safety 2}& \emph{Misleading information integrity}. The architecture must
  control the value of the LPV data provided by each calculator and
  between each screen and find mitigation in case of erroneous data.\\
\hline
\end{tabular}

In this paper we will first show how to model this architecture using
Alloy (Sect.~\ref{sec:eval-case-study}) and how to formally prove that
this architecture indeed complies to the aforementioned safety
requirements (Sect.~\ref{sec:assessm-safety-secur}).

In a second step, we want to ensure that the above architecture is
also resilient to a number of malevolent attacks
(Sect.~\ref{sec:security-assessment}). Seven attacks have been
considered as listed below.


\begin{tabular}{|l p{12.5cm}|}
\hline &\textbf{Security attacks:}\\ \emph{Attack 1}& One malicious
GPS signal (a fake signal that SBAS considers to come from
GPS).\\ 
\emph{Attack 2}& One constellation satellite signal is
scramble.\\ 
\emph{Attack 3}& The RNAV ground station is neutralized,
meaning that no more RNAV signal can be send to the
plane.\\ 
\emph{Attack 4}& An attack combining Attack 1 and Attack 2
scenarios.\\ 
\emph{Attack 5}& An attack combining Attack 2 and Attack
3 scenarios.\\ 
\emph{Attack 6}& An attack combining Attack 1 and
Attack 3 scenarios.\\ 
\emph{Attack 7}& An attack combining Attack 1,
Attack 2 and Attack 3 scenarios.\\ 
\hline
\end{tabular}

The paper shows how the proposed architecture fails to comply with all
the security requirements, and how the architecture can be enhanced to
provide full compliance to both safety and security requirements.



\section{Modelling the case study with Alloy}
\label{sec:eval-case-study}
Alloy~\cite{Jackson:06} is a formal system-modelling language amenable
to automatic analyses. Alloy is implemented as a cost-less
free-software tool, the Alloy Analyzer, which is programmed in Java
and hence runnable on the majority of platforms. Alloy has recently
been used in the context of security assessment, for instance to model
JVM security constraints~\cite{Reynold:10}, access control
policies~\cite{Toahchoodee:09}, or attacks in crytogrpahic
protocols~\cite{Lin:10}. Besides, we proposed in earlier work a
preliminary study of the safety assessment of the LPV
system~\cite{Brunel:14}. In this article, we propose to enrich this
study by considering a more complete architecture and by expressing
and verifying both safety and security objectives.

The AltaRica~\cite{Arnold:00} language, which is widely use for safety
assessment, would have been another possible choice.  However, we
decided to take benefit from the model-based aspect of Alloy and its
expressiveness for the specification of the properties to
check. Indeed, Alloy allows to define easily the metamodel of the
avionic architectures we will analyze instead of encoding them in
terms of AltaRica concepts. Moreover, the specification of the
properties we want to check are expressed in relational first-order
logic with many features adapted to model-based reasoning.

\subsection{General presentation of modelling in Alloy}
\label{sec:modelling-alloy}

Alloy is a general-purpose modelling language that is nonetheless very
well adapted to the following (non-exhaustive) list of activities:
abstract modelling of a problem or of a system; production of a
metamodel (model corresponding to a viewpoint); analysis of a model
using well-formedness or formal semantic rules; automatic generation
of an instance conforming to a model, possibly according to
supplementary constraints; finding interesting instances of a model.
Models designed in Alloy can deal with static aspects only, or
integrate also dynamic aspects, so as to check behavioral properties.

We now give a brief glance at the main concepts of
the language using a simple example. Each Alloy file contains one
module made of several declarations, the order of which is not
important. The most important declaration is the signature which
introduces a structural concept. This may be seen as a class or entity
in modelling parlance. A signature should be interpreted as a set of
possible instances (as it is sometimes said that a class “is” the set
of objects defined by itself). Finally, every signature comes with
fields that may be seen, as a first approximation, as class
attributes.

\begin{lstlisting}
sig Data { consumedBy : some System } 
sig System {} 
sig Criticality { 
	concernedData : one Data, 
	concernedSystem : one System 
}
\end{lstlisting} 

Here, we defined 3 concepts : Data, System and Criticality. As
explained earlier, Alloy advocates not to delve into unnecessary
details and only give information on things we want to understand or
analyze. Thus, here, a system is just defined to be a set of “things”,
but we do not say anything about the exact nature of its elements.

The keywords some or one give details on the multiplicity of the
relation, as 1..* and 1 in UML. Here the field declarations mean that:
every datum is consumed by at least one (some) system; every
criticality concerns exactly one (one) data and one system.  Other
possible multiplicities are: lone which means at most one; and set
which means any number (including 0). 

Then, we can add constraints on possible instances of our model. For
instance, we would like to state that every system consumes at least
one datum. This can be done by writing additional facts (facts state
properties that always hold, so as few facts as possible should be
stated in order to avoid over-specification):

\begin{lstlisting}
fact { 
	// every system consumes at least one datum 
	all s : System | some consumedBy.s
	// for any system which consumes a given datum, the said datum and system 
	// should belong to a same unique criticality 
	all d : Data | all s : System | one c : Criticality | 
		c.concernedData = d and c.oncernedSystem = s 
} 
\end{lstlisting}


Given all these definitions, the Alloy Analyzer can 
\begin{itemize}
\item provide graphical representation of the metamodel or of an instance
\item carry out some explorations (the command \lstinline{run} builds instances that satisfy a given statement)
\item check whether a given constraint is satisfied by all instances of the model (command \lstinline{check})
\end{itemize}





\subsection{LPV model in Alloy}
\label{sec:lpv-model-alloy}
\begin{figure}[hbtp]
\centering
  \includegraphics[width=.99\textwidth]{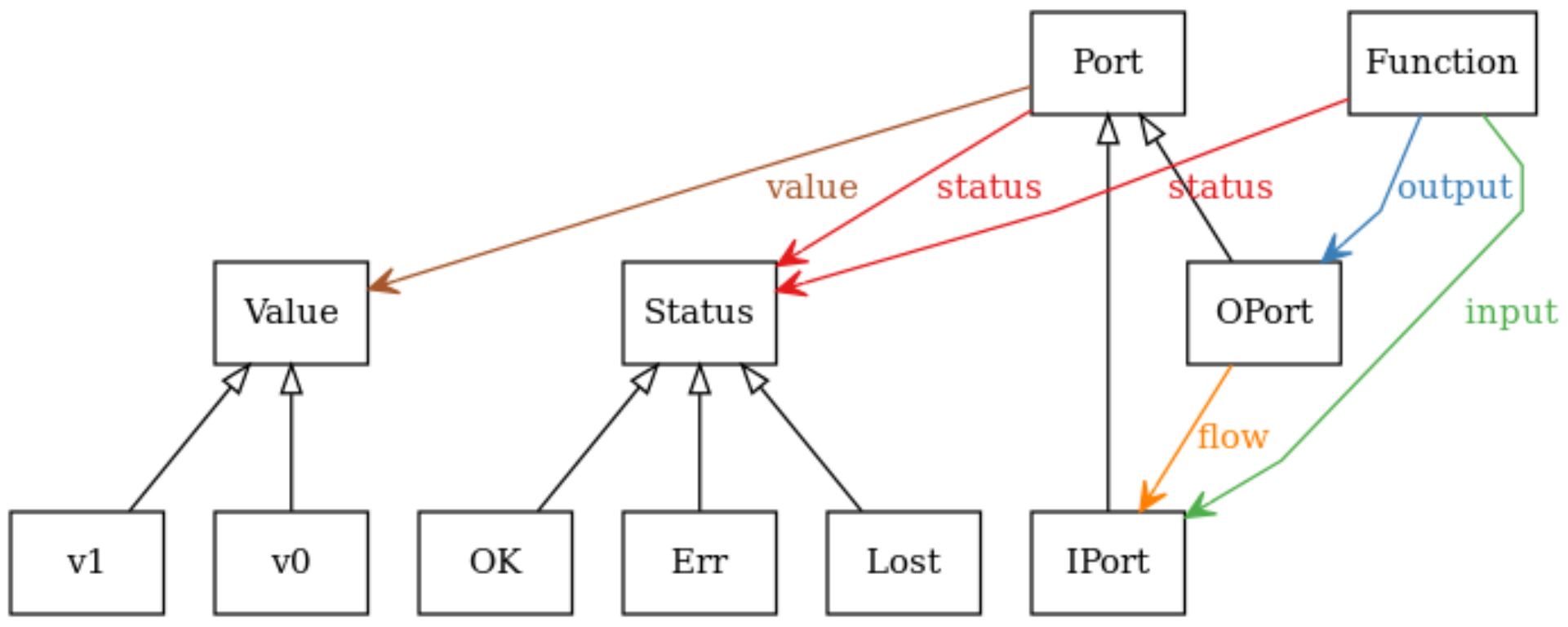}
  \caption{Case study metamodel (simplified)}
\label{fig:mm}
\end{figure}

To represent the functional architecture of LPV (see
Fig~\ref{fig:LPV-archi}) in Alloy, we identified the major concepts at
stake and defined an Alloy signature for each of them:
\lstinline{Function}, \lstinline{Port}, \lstinline{IPort} (input
port), \lstinline{OPort} (output port). Each Function has a set
\lstinline{input} of \lstinline{IPorts} and a set \lstinline{output}
of \lstinline{OPort} as attributes. An \lstinline{OPort} of a function
is related to a set of \lstinline{IPort} of other functions through a
\lstinline{flow}. For a given function, the relation between its
\lstinline{IPort} and \lstinline{OPort}, for instance expressing how
failures propagate, can be given directly by an Alloy formula (a
\emph{fact} in Alloy terminology) that constrains the model. Each
function and each port hold a dysfunctional \lstinline{status}
(\lstinline{OK}, \lstinline{Lost}, or \lstinline{Err}). If the status
of a port is \lstinline{Lost}, it means that this port does not
produce any data. If the status is \lstinline{Err}, it means that it
yields an undetected erroneous data. Moreover, each port has a
\lstinline{value}. Notice that, in our model, the value will only be
useful for certain ports, representing the pilot selection, the
discrepancy between both LPV processing, and the reset of
displays. Concerning the other ports, we only deal with their status
and the value is just ignored (a finer modelling would have been
possible here but it would have led to a worthless, more complex
model). The following Alloy code corresponds to these concept
definitions (see also Fig.~\ref{fig:mm}).

\begin{lstlisting}
enum Status { OK, Err, Lost }
abstract sig Port {
         status : Status,
         value : Value
}
abstract sig IPort extends Port {}
abstract sig OPort extends Port {
         flow : set IPort,
}
abstract sig Function {
         input : set IPort,
         output : set OPort,
         status : Status
} 
\end{lstlisting}

We then define instances of these concepts corresponding to the LPV
functional architecture. The function instances take into account the
selection of the source by the crew (\lstinline{SelectSource}), the
satellite data position (\lstinline{GPS} and \lstinline{Galileo}), two
occurrences of SBAS positioning (\lstinline{ComputeSBAS1},
\lstinline{ComputeSBAS2}), two occurrences of LPV processing
(\lstinline{ComputeLPV1}, \lstinline{ComputeLPV2}), three occurrences
of displays (\lstinline{Acquirei}, $i\in \{1..3\}$), three occurrences
of display resetters (\lstinline{Crosschecki}, $i\in \{1..3\}$) and of
monitors in order to trigger an alarm, (\lstinline{Monitori}, $i\in
\{1..3\}$). We also define the different ports of each function, and
the way ports are related to each other via flows.  For instance, the
following Alloy code is an excerpt of the flow definition, expressing
that the output port \lstinline{oSBAS1} is related to the input port
\lstinline{iSBAS1} via a flow (idem for \lstinline{oSBAS2} and
\lstinline{iSBAS2}).
	
\begin{lstlisting}
flow = oSBAS1->iSBAS1 + oSBAS2->iSBAS2 +  ...
\end{lstlisting}

We also define some global constraints the architecture must satisfy,
such as the fact that two ports related by a flow share the same status and
the same value:
\begin{lstlisting}
all p1, p2 : Port | p1->p2 in flow implies p1.status = p2.status and p1.value = p2.value
\end{lstlisting}

We now define the relation between input and output ports inside each
function in term of failure propagation. For instance, the following
code expresses that the status of the output port \lstinline{oDeviation1} of
function \lstinline{ComputeLPV1} is equal to the status of the input if the
function \lstinline{ComputeLPV1} is OK, is equal to Lost if the
function \lstinline{ComputeLPV1} is lost, and is erroneous (\lstinline{Err})
otherwise.
\begin{lstlisting}
let st = ComputeLPV1.status |
oDeviation1.status = {
                   st = OK implies iSBAS1.status
                   else st = Lost implies Lost
                   else Err 
}
\end{lstlisting}

The following code defines the status of the output port \lstinline{oSelected1}
of function \lstinline{Aquire1}. If this function is OK, \lstinline{oSelected1}
the status is either equal to the status of its first input or to the status of
its second input, depending on the selection made by the pilot. If the function
\lstinline{Aquire1} is lost, then the status of output \lstinline{oSelected1} is
lost. Otherwise, it is erroneous.

\begin{lstlisting}
let st = Acquire1.status | 
let v = iSelection1.value | 
oSelected1.status = { 
                  st = OK and v = v0 implies iDeviation11.status
                  else st = OK and v = v1 implies iDeviation21.status
                  else st = Lost implies Lost
                  else Err 
}
\end{lstlisting}

Similarly, we define, for each function, output ports status (and
value in the case of pilot selection, discrepancy, and display reset)
from input ports status (and value).

\begin{figure}[h!]
\begin{center}
\includegraphics[width=0.85 \textwidth]{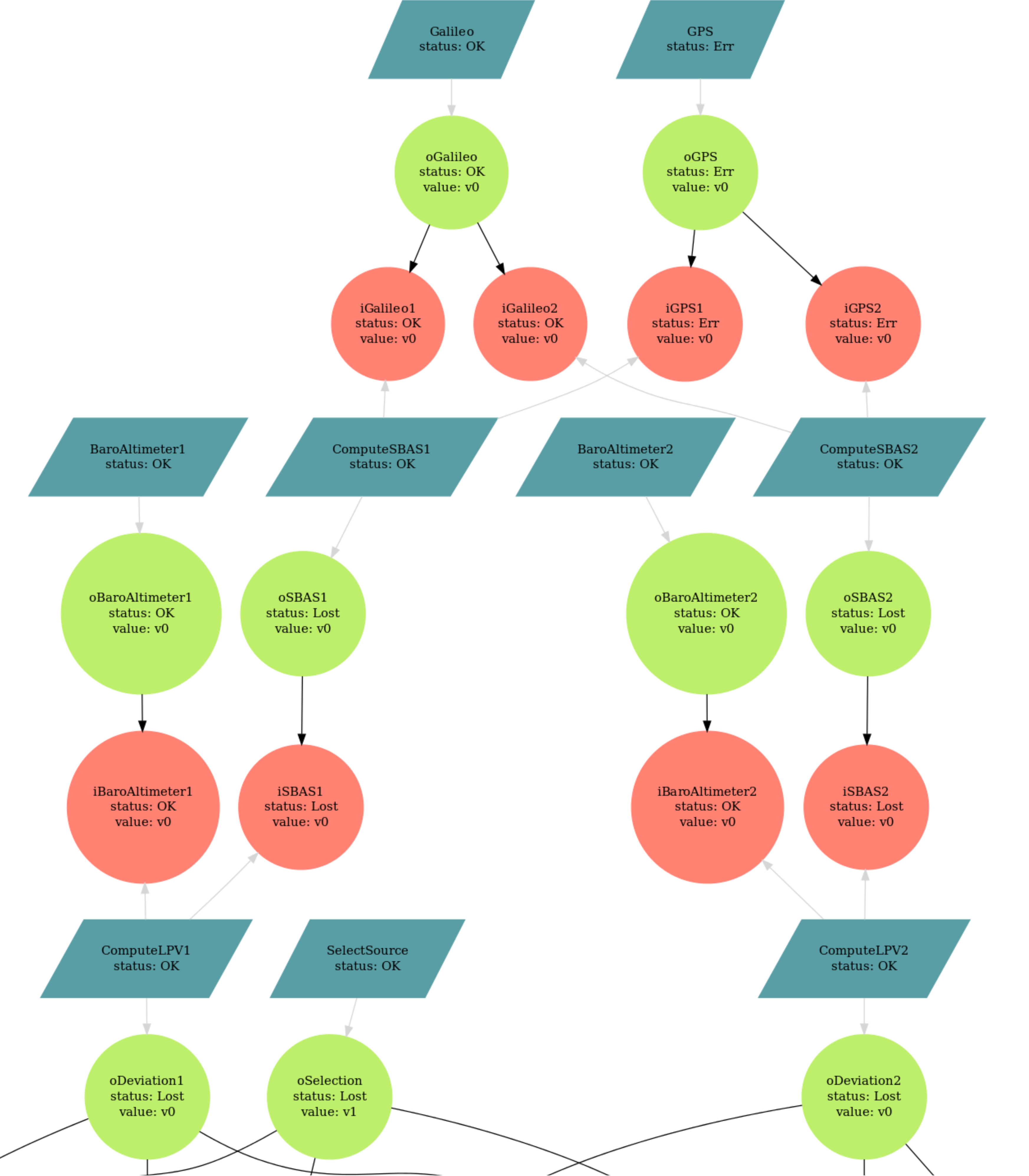}
\caption{\label{fig:fun-archi-alloy}Visualization of LPV functional architecture in the Alloy Analyzer (excerpt)} 
\end{center}
\end{figure}

\bigskip

Then, Alloy Analyzer proposes, with the command \lstinline{run} to produce
automatically an instance that satisfies all the definitions we expressed, and
the global constraints. Of course, this is only possible if there is no
inconsistency in the definitions. We can customize the way the instance is
graphically displayed. Fig.~\ref{fig:fun-archi-alloy} shows an excerpt of an
instance produced by the Analyzer, where functions are represented by blue
parallelograms, output ports by green circles and input ports by red
circles.

\bigskip

\section{Safety assessment}
\label{sec:assessm-safety-secur}
We are now able to check formally various properties expressed in
Alloy. A first kind of properties we can check consists of structural
properties about the model. For instance, the fact that a port can
only belong to one function and the fact that flows correspond to
one-to-many communication is expressed by the following properties:

\begin{lstlisting}
assert model_structure {
	input in Function one -> IPort
	output in Function one -> OPort
	flow in OPort one -> IPort }
\end{lstlisting}

The command \lstinline{check model_structure} verifies that, \emph{up to a
certain bound}, all possible instances of the model satisfy these properties. If
it is not the case, it yields a counter-example instance.
\pagebreak

Now we want to validate the safety objectives expressed in Sect.~\ref{sec:LPV}.
Concerning the \emph{Loss of LPV capability} constraint, we express the
following property:  \medskip

\noindent$\bullet$ \emph{If one (and only one) LPV processing is lost, and if
the pilot makes a correct selection, then the data sent by the three displays are
still correct.} This corresponds to the following Alloy code (for the loss of
LPV1):

\begin{lstlisting}[frame=single]
assert one_computer_lost {
	(all f: Function | (f != ComputeLPV1 implies f.status=OK) 
                     and ComputeLPV1.status=Lost and oSelection.value=v1)
	implies oSelected1.status = OK 
        and oSelected2.status = OK
        and oSelected3.status = OK 
}
\end{lstlisting}

The command \lstinline{check one_computer_lost} verifies that the model satisfies this property.

\medskip Concerning the \emph{Misleading information integrity}
constraint, we expressed two properties to be satisfied by the
functional architecture:

\noindent$\bullet$ \emph{If one LPV processing produces erroneous data, then an
alarm (modeled by the variables oDiscrepancy) is launched on the three
displays.} This corresponds to the following Alloy code (for the LPV1 in
erroneous failure mode):
\begin{lstlisting}[frame=single]
assert one_computer_erroneous {
	(all f: Function | (f != ComputeLPV1 implies f.status=OK) 
                     and ComputeLPV1.status=Err)
  implies oDiscrepancy1.value=v1
        and oDiscrepancy2.value=v1
        and oDiscrepancy3.value=v1 
}
\end{lstlisting}

\medskip

\noindent$\bullet$ \emph{If one display returns an erroneous data,
it resets itself.} This corresponds to the following Alloy code for display 1
(\lstinline{Acquire1}) in erroneous failure mode:
\begin{lstlisting}[frame=single]]
assert one_display_erroneous {
  (all f: Function | (f != Acquire1 implies f.status=OK) 
                     and  Acquire1.status=Err)
  implies oReset1.value=v1 
}
\end{lstlisting}

These safety objectives are validated by the Alloy Analyzer.


\section{Security assessment}
\label{sec:security-assessment}
Let us now consider the security objectives. For each attack
identified in Sect.~\ref{sec:LPV}, we will express that the attack has
no bad influence on the LPV capability, \emph{i.e.}, the data produced
by the three displays (represented by ports \lstinline{oSelectedi})
are correct. So, we have the following expression for the first two
identified attacks.

{\small 
\begin{lstlisting}[frame=single]
//Attack1
assert one_satellite_corrupted {
	(all f: Function | (f != GPS implies f.status=OK) and GPS.status=Err)
	implies oSelected1.status = OK and oSelected2.status = OK and oSelected3.status = OK
}
//Attack2
assert one_satellite_lost{
	(all f: Function | (f != GPS implies f.status=OK) and GPS.status=Lost)
	implies oSelected1.status = OK and oSelected2.status = OK and oSelected3.status = OK
}
\end{lstlisting}
}

It turns out that one of these properties is false with the current
version of the model, which does not include RNAV function for lateral
guidance. Alloy Analyzer produces counter examples that help
understanding why the LPV capability is lost. Actually, without RNAV,
the loss of SBAS satellite data is sufficient to lose the whole LPV
capability. After the first attack, \emph{i.e.}, a fake GPS signal,
the SBAS does not manage to consolidate both satellite data, so the
LPV processing does not have any input data, which induces a loss of
the signal that is sent to the displays. However, the second attack,
which consists in scrambling one satellite data (the corresponding
signal is lost) is not sufficient to lose the LPV capability since
SBAS sends to LPV processing the other satellite data.

In order to resist to the first attack, we propose to consider, in
addition to satellite data, RNAV signal together with a
Baro-altimeter, and to provide an alarm signal to the pilot when the
difference between the information of SBAS and of the couple
RNAV-Baro-altimeter is too important. 

The definition of the output of LPV processings becomes more
complicated, since it takes into account SBAS data, RNAV and
baro-altimeter. Besides, it provides an alarm to the pilot in case of
inconsistency between SBAS data on the one hand and
RNAV/baro-altimeter data on the other hand. These definitions for the
first LPV processing (the second is similar) are given as follows.

{\small 
\begin{lstlisting}[frame=single]
let st = ComputeLPV1.status | oDeviation1.status = {
  st=Lost implies Lost else st=Err implies Err else iSBAS1.status = OK implies OK 
  else iSBAS1.status = Lost and iRNAV1.status=OK and iBaroAltimeter1.status=OK 
       implies OK 
  else iSBAS1.status=Lost and (iRNAV1.status=Lost or iBaroAltimeter1.status=Lost) 
       implies Lost
  else iSBAS1.status=Lost and (iRNAV1.status=Err or iBaroAltimeter1.status=Err) implies Err 
  else iSBAS1.status=Err and (iRNAV1.status=Lost or iBaroAltimeter1.status=Lost) implies Err 
  else iSBAS1.status=Err and iRNAV1.status=OK and iBaroAltimeter1.status=OK implies Lost 
  else iSBAS1.status=Err and (iRNAV1.status=Err or iBaroAltimeter1.status=Err) implies Lost 
  else Err } 

let st = ComputeLPV1.status | LPV1_alarm.value = { 
  st=OK and iSBAS1.status != Lost and iRNAV1.status != Lost and iBaroAltimeter1.status != Lost
  and (iSBAS1.status=Err or iRNAV1.status=Err or iBaroAltimeter1.status=Err) 
  implies v1 
  else st=OK and iSBAS1.status= Err and (iRNAV1.status=OK or iBaroAltimeter1.status=OK)
       implies v1
  else st=OK and (iSBAS1.status = Lost) 
       and (iRNAV1.status=Lost or iBaroAltimeter1.status=Lost) implies v1 
  else v0 }
\end{lstlisting}
}

Now we can define the security objectives corresponding to all the
attacks identified in Sect.~\ref{sec:LPV}.
 
{\small \begin{lstlisting}[frame=single]
//Attack 3 
assert  RNAV_lost {
	(all f: Function | (f != RNAV1 and f != RNAV2 implies f.status=OK) 
        and RNAV1.status=Lost and RNAV2.status=Lost)
	implies oSelected1.status = OK and oSelected2.status = OK and oSelected3.status = OK
}
//Attack 4
assert one_satellite_lost_one_satellite_corrupted {
	(all f: Function | (f != GPS and f != Galileo) implies f.status=OK 
        and GPS.status=Err and Galileo.status=Lost)
	implies LPV1_alarm.value=v1
}
//Attack 5
assert one_satellite_lost_RNAV_lost {
	(all f: Function | (f != GPS and f != RNAV1) implies f.status=OK)
        and GPS.status=Lost and RNAV1.status=Lost
	implies oSelected1.status = OK and oSelected2.status = OK and oSelected3.status = OK
}

//Attack 6
assert one_satellite_corrupted_RNAV_lost {
	(all f: Function | (f != GPS and f != RNAV1 and f !=RNAV2) implies f.status=OK) 
        and GPS.status=Err and RNAV1.status=Lost and RNAV2.status=Lost
	implies LPV1_alarm.value=v1
}

//Attack 7
assert one_satellite_corrupted_one_satellite_lost_RNAV_lost  {
       (all f: Function | f != GPS and f != Galileo and  f != RNAV1 and f !=RNAV2 
          implies f.status=OK) 
       and GPS.status=Err and Galileo.status=Lost 
       and RNAV1.status=Lost and RNAV2.status=Lost
       implies LPV1_alarm.value=v1
}
\end{lstlisting}
}

Notice that for properties relative to attacks 4, 6 and 7, we only
require the alarm to be launched (the information that come to the
displays are incorrect or absent in these cases). For the other
properties, we require that the information displayed are correct.

All these properties are validated by the Alloy Analyzer.

\section{Conclusion}

In this article, we modeled the architecture of an avionic system in
the Alloy language. We showed how Alloy allows to easily define a
simple metamodel of avionic architectures with failure propagation,
which particularly fits the representation of the architecture of our
case study.

We then expressed safety and security properties and enriched the
initial model in order to fulfill all the objectives. We checked with
the Alloy Analyzer that these properties are fulfilled by the model.

Since the safety and the security properties that we expressed are
based on the same failure conditions, it would be possible to mix the
assessment of both concerns, in order for instance to study the impact
of attacks on safety objectives, or the other way around, the impact
of unintentional failures on security objective. Such kinds of
analyses of safety and security together, for which lightweight formal
methods and Alloy in particular are promising, will be studied in
particular in the remainder of the MERgE project.


\bibliographystyle{eptcs}
 \bibliography{main}
\end{document}